\documentclass[a4paper]{article}
\usepackage{INTERSPEECH2019}
\usepackage{amsmath,graphicx}
\usepackage{multirow}
\usepackage{mathtools,xparse}
\usepackage{graphicx}
\usepackage[dvips]{epsfig}
\usepackage{adjustbox}
\usepackage{subcaption}
\usepackage{cite}
\usepackage{array}
\usepackage{makecell}
\usepackage{multicol}
\usepackage{amsmath}
\usepackage{mathabx}
\usepackage{caption}
\usepackage{algorithm}
\usepackage{algpseudocode}
\usepackage{tabularx}
\usepackage{float}
\PassOptionsToPackage{hyphens}{url}\usepackage{hyperref}

\title{Effective Parameter Estimation Methods for an ExcitNet Model \\in Generative Text-to-Speech Systems}
\name{{Ohsung Kwon$^{1,2,*}$, Eunwoo Song$^{1}$, Jae-Min Kim$^1$ and Hong-Goo Kang$^{2}$}
\address{ $^1$NAVER Corp., Seongnam, Korea \\
$^2$Department of Electrical and Electronic Engineering, Yonsei University, Seoul, Korea}
\thanks{*Work partially performed as an intern in Clova Voice, Naver Corp., Seongnam, Korea.}
\email{osungv@dsp.yonsei.ac.kr, \{eunwoo.song, kjm.kim\}@navercorp.com, hgkang@yonsei.ac.kr}
}

\begin{document}
\maketitle
\begin{abstract}

    In this paper, we propose a high-quality generative text-to-speech (TTS) system using an effective spectrum and excitation estimation method.
    Our previous research verified the effectiveness of the ExcitNet-based speech generation model in a parametric TTS framework.
    However, the challenge remains to build a high-quality speech synthesis system because auxiliary conditional features estimated by a simple deep neural network often contain large prediction errors, and the errors are inevitably propagated throughout the autoregressive generation process of the ExcitNet vocoder.
    To generate more natural speech signals, we exploited a sequence-to-sequence (seq2seq) acoustic model with an attention-based generative network (e.g., Tacotron 2) to estimate the condition parameters of the ExcitNet vocoder.
    Because the seq2seq acoustic model accurately estimates spectral parameters, and because the ExcitNet model effectively generates the corresponding time-domain excitation signals, combining these two models can synthesize natural speech signals.
    Furthermore, we verified the merit of the proposed method in producing expressive speech segments by adopting a global style token-based emotion embedding method.
    The experimental results confirmed that the proposed system significantly outperforms the systems with a similarly configured conventional WaveNet vocoder and our best prior parametric TTS counterpart.

\end{abstract}
\vspace*{-4pt}

\section{Introduction}
\label{sec:intro}

    Waveform generation systems using WaveNet have attracted a great deal of attention in speech-signal processing communities thanks to their high quality and ease of use in various applications \cite{van2016conditional, van2016wavenet}.
    In these types of systems, the time-domain speech signal is represented as a sequence of discrete symbols, and its distribution is autoregressively modeled by stacked convolutional layers.
    By appropriately conditioning the acoustic features to the input, WaveNet-based systems have also been successfully applied to a neural vocoder structure for parametric text-to-speech (TTS) and end-to-end generative TTS systems \cite{tamamori2017speaker, hayashi2017investigation, adiga2018use, wang2017tacotron, Shen2018NaturalTS, Ping2017DeepV3, Ping2018ClariNetPW}.
    
    To further improve the perceptual quality of synthesized speech, more recent neural excitation vocoders (e.g., ExcitNet \cite{song2018excitnet}) take advantage of the merits from both a linear prediction (LP)-based parametric vocoder and the WaveNet structure \cite{yoshimura2018mel, juvela2018speaker, valin2018lpcnet, wang2018neural, tachibana2018investigation}.
    In this framework, an adaptive predictor is used to decouple the formant-related spectral structure from the input speech signal, and then the probability distribution of its residual signal (i.e., the excitation signal) is modeled by the WaveNet network.
    As variation in the excitation signal is only constrained by vocal cord movement, the training and generation processes become much more efficient. 

    However, this approach still requires an accurate acoustic model to enable effective conversion of a text sequence into corresponding acoustic parameters.
    Since these estimated parameters are directly used as an input of the ExcitNet, prediction errors can be boosted throughout the autoregressive process for producing an excitation signal. 
    Furthermore, estimated spectral parameters are used to transform generated excitation signals into speech signals; thus, accurately estimating these parameters is crucial to reconstruct natural speech signals.

    For this paper, we exploited a sequence-to-sequence (seq2seq) acoustic model with an attention-based generative network (e.g., Tacotron 2) to estimate acoustic parameters for an ExcitNet-based generative TTS system \cite{wang2017tacotron, Shen2018NaturalTS}. 
    It is well known that Tacotrons can effectively produce a magnitude spectrogram from a sequence of input text strings by replacing the complicated pipeline of conventional acoustic models with a single neural network; therefore, it has demonstrated better quality than the conventional parametric TTS applications have.
    The proposed method adopted a similar structure to the Tacotron 2, but the model was guided to estimate the spectral and excitation parameters needed for the ExcitNet framework. 
    First, the input text sequence was encoded into transcript embedding by the seq2seq transcript encoder, and then the embedding was passed through a location-sensitive attention network for alignment with the target acoustic parameters \cite{chorowski2015attention}.
    After that, the seq2seq decoder recursively predicted the target acoustic parameters followed by the ExcitNet-based excitation generator. 
    Finally, a speech signal was synthesized by the generated spectral parameters and the excitation signal.

    Using the baseline system explained above, we also developed a global style token (GST)-based expressive TTS system. 
    Note that the GST is a high-dimensional embedding that implicitly contains the speaker's prosody and style information \cite{wang2018style}.
    In our previous work, we verified the superiority of the GST-based emotional speech synthesis model over the conventional approaches \cite{kwon2019emo}. 
    Furthermore, as the ExcitNet model has performed robustly in producing prosody-modified speech segments \cite{song2018speaker}, combining the GST, Tacotron 2, and ExcitNet models enables significant improvement in the perceptual quality of the synthesized speech.

    We investigated the effectiveness of the proposed method by conducting subjective evaluation tasks.
    The experimental results show that the proposed speech synthesis system provides significantly better perceptual quality than the conventional methods designed with a similarly configured WaveNet vocoder and our best prior parametric TTS counterpart.

    \begin{figure*}[t!]
        \centering
        \includegraphics[width=\textwidth]{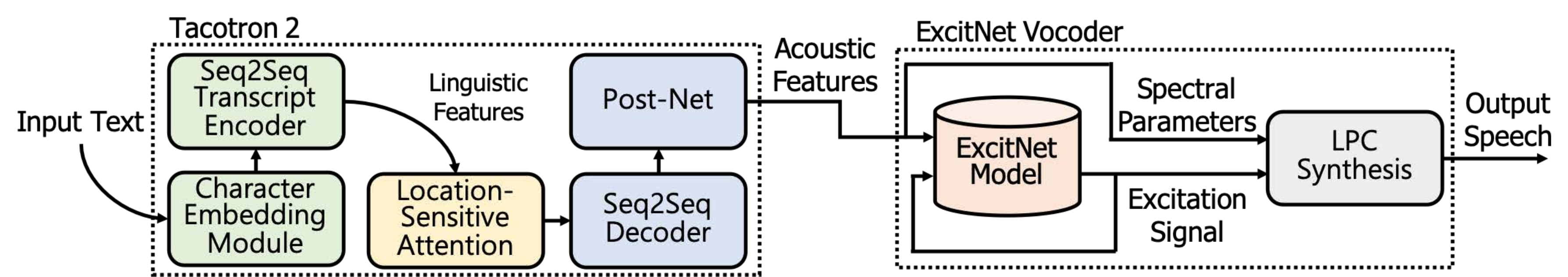}
        \vskip -3pt
        \caption{Proposed speech synthesis framework based on Tacotron 2 with ExcitNet vocoder.}
        \label{fig:block}
        \vskip -7pt
    \end{figure*}

\section{Relationship to prior work} 
\label{sec2}

    The idea of using a neural ExcitNet model for a speech synthesis system is not very new.
    By incorporating an LP filter into the WaveNet architecture, the ExcitNet model significantly improves the generative accuracy of the vocoder \cite{song2018excitnet}. 
    However, the quality of the synthesized speech is often degraded when the ExcitNet vocoder is combined with the front-end parameter estimation network; in other words, if the input conditional features of the ExcitNet contain large estimation errors, the generated speech often includes unexpected artifacts.
    In addition, the estimated spectral parameters to be used for the LP synthesis filter are too smooth; thus, the synthesized speech becomes muffled.
    
    To ameliorate these issues, we replaced the front-end parameter estimation module with a seq2seq model with an attention-based approach (e.g., similar to the front-end of Tacotron 2) to estimate acoustic parameters accurately.
    The proposed approach differs from Tacotron 2 because the estimated output parameters are designed for the ExcitNet architecture in the following stage.
    Note that the conventional seq2seq models for generative TTS applications have only been combined with the WaveNet architecture \cite{Shen2018NaturalTS, Ping2017DeepV3, Ping2018ClariNetPW}.
    Furthermore, we explored the use of the GST-Tacotron 2 with the ExcitNet vocoder to present expressiveness in the synthesized speech.
    Analysis shows that the proposed method significantly improves the naturalness of synthesized speech in both narrative and expressive speech synthesis tasks when compared with the conventional methods.
    In addition, regarding the vocoder itself, in a perceptual listening test, the proposed system is superior to both our best prior parametric improved time-frequency trajectory excitation (ITFTE) vocoder and the conventional WaveNet vocoder with the same seq2seq model structure.

\section{ExcitNet-based neural vocoding model}
\label{sec:ch3}

\subsection{WaveNet vocoder}
\label{sec:ch3-1}

    The basic WaveNet uses autoregression to generate a probability distribution of the waveform from a fixed number of past samples, and additional auxiliary features can control the characteristic of the output signal. 
    Typically, linguistic features, the fundamental frequency (F0), and/or speaker codes are used for the auxiliary features \cite{van2016wavenet}.

    More recent \textit{WaveNet vocoders} utilize another form of acoustic parameters directly extracted from input speech, such as mel-filterbank energy, mel-generalized cepstrum, band aperiodicity, and F0 \cite{tamamori2017speaker, hayashi2017investigation, adiga2018use}. 
    Since the acoustic parameters include relevant information that enables automatic learning of the relationship to speech samples, the approaches could obtain superior perceptual quality over traditional parametric vocoding-based approaches \cite{tamamori2017speaker, wang2018comparison}.
    
\subsection{ExcitNet vocoder}
\label{sec:ch3-2}

    Even though previous studies have indicated the technical feasibility of using WaveNet for a vocoding process, the systems often have noisy outputs; this is due to inherent structural limitations when capturing the dynamic nature of speech signals caused by the convolutional neural network structure \cite{tachibana2018investigation}.

    To improve the perceptual quality of synthesized speech, several frequency-dependent noise-shaping filters have been proposed \cite{song2018excitnet, tachibana2018investigation, yoshimura2018mel, juvela2018speaker, valin2018lpcnet, wang2018neural}.
    In particular, the neural excitation vocoder \textit{ExcitNet} described in the right-hand part of Figure~\ref{fig:block} exploits an LP-based adaptive predictor to decouple the spectral formant structure from the input speech signal.
    The WaveNet-based generation model is then used to train the residual LP component (i.e., the excitation signal).
    As vocal cord movement is the only constraint on variation in the excitation signal, the training process becomes much more effective.

    In the speech synthesis step, the spectral and excitation features of the given input are first generated by a deep learning-based acoustic model.
    Then, these parameters are used as input conditional features for the ExcitNet model to generate the corresponding time sequence of the excitation signal. 
    Finally, a speech signal is reconstructed by passing the generated excitation signal through the spectral synthesis filter. 
    
    As the spectral parameters are used not only to compose the input of the ExcitNet model but also to transform the generated excitation signal into the speech domain, accurately estimating these parameters is crucial to generate natural speech signals. 
    For this paper, we exploited the state-of-the-art generative TTS model (i.e., Tacotron 2) as an acoustic model to further improve the accuracy of the feature estimation process.

\section{Seq2seq acoustic model}
\label{sec:ch4}

\subsection{Tacotron 2}
\label{sec:ch4-1}

    The left-hand part of Figure~\ref{fig:block} represents the \textit{Tacotron 2}-based acoustic model \cite{Shen2018NaturalTS}, where the seq2seq-based generative network maps the input text to the corresponding acoustic features consisting of spectral and excitation parameters.
    There are three sub-networks: a seq2seq transcript encoder, a location-sensitive attention network, and the seq2seq decoder. 
    The seq2seq transcript encoder first converts the input text sequence (phoneme- or grapheme-level) into the same-level transcript embedding as hidden linguistic features, and then they are transferred to the seq2seq decoder via the attention network in every decoding step.
    Note that the decoder comprises the recursive network, which estimates frame-level acoustic features from the hidden linguistic features, and the previously predicted acoustic features.  
    Because the three sub-networks act as a single generative network, acoustic features are more accurately generated when compared to the conventional acoustic parameter estimation model.
    
\subsection{GST-based emotional style embedding}
\label{sec:ch4-2}

    \begin{figure}[t!]
        \centering
        \includegraphics[width=0.52\textwidth]{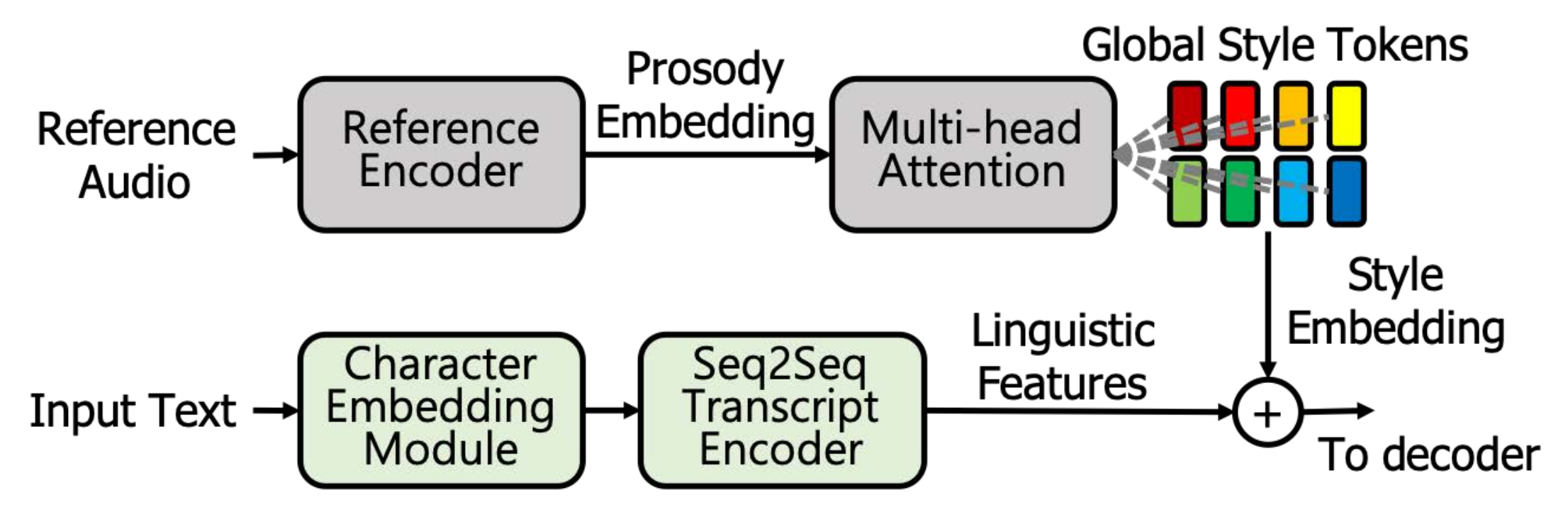}
        \vskip -3pt
        \caption{A block diagram of the GST-based style modeling network.}
        \label{fig:gst}
        \vskip -7pt
    \end{figure}

    When the Tacotron 2 model is combined with the GST-based style embedding method, expressiveness can be given to the synthesized speech segments \cite{wang2018style}.
    Note that the GST involves high-dimensional embedding that implicitly contains the speaker's prosody and style information. 
    As shown in Figure~\ref{fig:gst}, a reference encoder and a style token layer, which contain an additional multi-head attention network with a bank of GSTs, are incorporated into the Tacotron 2 model to generate the style embedding, which is conditioned by the transcript embedding.
    The reference encoder extracts prosody embedding from the input reference audio, and the attention network measures similarities between the prosody embedding and each GST.
    Then, weights extracted from the attention network are linearly combined with the GSTs to generate style embedding.

    Because the GSTs can deal with a highly dynamic range of expressive speech, the style embedding is helpful to generate more expressive speech signals, and it can be adapted for emotional speech synthesis \cite{kwon2019emo}.
    Furthermore, as the ExcitNet vocoder is effective at generating prosody-modified speech segments \cite{song2018speaker}, speech signals generated by combining these two models were expected to be more natural than those generated by conventional approaches.
    
\section{Experiments}
\label{sec:ch5}

\subsection{Tacotron 2 with ExcitNet vocoder}
\label{ssec:ch5-1}
\subsubsection{Database}
\label{ssec:ch5-1-1}

    Two phonetically and prosodically rich speech corpora were used to train the acoustic model and the ExcitNet vocoder.
	Each corpus was recorded by professional Korean female and Korean male speakers.
	The speech signals were sampled at 24 kHz, and each sample was quantized by 16 bits.
	Table~\ref{table:numUtt} shows the number of utterances in each set.   

    To compose the acoustic feature vectors, the spectral and excitation parameters were extracted using a previously proposed parametric ITFTE vocoder \cite{song2017effective}.
    In this way, 40-dimensional line spectral frequencies (LSF), 32-dimensional slowly evolving waveform (SEW), 4-dimensional rapidly evolving waveform (REW), the F0, the gain, and the v\texttt{/}uv were extracted. 
    The frame and shift lengths were set to 20 ms and 5 ms, respectively.

\subsubsection{ExcitNet-based neural vocoding model}
\label{ssec:ch5-1-2}    

    In the ExcitNet training step, all acoustic feature vectors were duplicated from a frame to the samples to match the length of the input speech signals \cite{song2018excitnet}. 
    Before training, they were normalized to have zero mean and unit variance. 
    The corresponding excitation signal was obtained by passing the speech signal through an LP analysis filter, then normalized to have a range between -1.0 and 1.0, and finally encoded by 8-bit $\mu$-law compression.
    The model architecture comprised three convolutional blocks, each with 10 dilated convolution layers with dilations of 1, 2, 4, and so on, up to 512. 
    The number of channels of dilated causal convolution and the 1$\times$1 convolution in the residual block were both set to 512. 
    The number of 1$\times$1 convolution channels between the skip-connection and the softmax layer was set to 256.
    The learning rate was set to 0.0001, and the batch size was set to 30,000 (1.25 sec).

    The setups for training a baseline WaveNet vocoder were the same as those for the ExcitNet, but the WaveNet predicted the distribution of the speech signal, noise-shaped by a time-invariant spectral filter \cite{tachibana2018investigation}.
    Note that the noise-shaping filter was obtained by averaging all spectra extracted from the training data.
    This filter was used to extract the residual signal before the training process, and its inverse filter was applied to reconstruct the speech signal in the synthesis step. 

	\begin{table}[!t]   
	\begin{center}         
	\caption{Number of utterances in different sets for the Korean male (KRM) and the Korean female (KRF) speakers (SPKs).}  
	\vspace*{-8pt}
	\label{table:numUtt}
	{\small        
	\begin{tabular}{>{\centering}m{.17\linewidth}||c|c|c}
	\Xhline{2\arrayrulewidth}
	SPK			& Training  & Development 	& Test \\
			\hline \hline
	KRF		& 4,976 (10.0 h)		& 280 (0.5 h)			& 280 (0.5 h)	\\
	KRM 	& 3,300 (10.0 h)		& 330 (0.5 h) 			& 330 (0.5 h)	\\
			\Xhline{2\arrayrulewidth}
	\end{tabular}}          
	\end{center}         
	\vspace*{-12pt}
	\end{table}

\subsubsection{Seq2seq acoustic model}
\label{ssec:ch5-1-3}

    In the Tacotron 2-based acoustic model training, input text and target acoustic feature pairs were prepared.
    Specifically, target acoustic features were normalized to have zero mean and unit variance before training.
    The corresponding input text sequence was converted to 512-dimensional embeddings by a front-end character module, and then they were fed into the seq2seq transcript encoder to extract transcript embedding.
    The transcript encoder had three convolution layers with 10$\times$1 kernel and 512 channels per layer, where the final layer was followed by a bi-directional long short-term memory (LSTM) layer with a total of 512 memory blocks.

    To align the transcript embedding with the target acoustic parameters, the location features were extracted by the location-sensitive attention network, the convolution layers of which comprising 63$\times$1 kernel and 64 channels.
    Those location features were passed through a fully connected layer with 128 units to be projected, which were utilized for a scoring mechanism to generate the contextual vector.

    By using the contextual vector, the seq2seq decoder estimated the acoustic features.
    First, previously generated acoustic features were fed into two fully connected layers with 256 units in each layer to extract bottleneck features. 
    Second, both contextual vectors from the attention network and the bottleneck features were passed through two uni-directional LSTM layers with 1,024 units, where the LSTM layer was followed by two projection layers generating the stop token and acoustic features.
    Finally, to improve generation accuracy, five convolution layers, with 5$\times$1 kernel and 512 channels per layer, were introduced as a post-processing network (i.e., post-net) to add the residual element of the generated acoustic features.
    
    During training, the mean squared error between the target and predicted acoustic features was used for the optimization criterion. 
    The learning rate was scheduled for decay from 0.001 to 0.0001 via a decaying rate of 0.33 per 100,000 steps.

    To construct a baseline parametric TTS framework \cite{song2017effective}, the output feature vectors comprised all the acoustic parameters and their time dynamics \cite{furui1986speaker}.
    The linguistic input feature vectors were 356-dimensional contextual information comprising 330 binary features of categorical linguistic contexts and 26 features of numerical linguistic contexts. 
	Before training, both input and output features were normalized to have zero mean and unit variance.
    The hidden layers comprised three feed-forward layers with 1,024 units and a uni-directional LSTM layer with 512 memory blocks.
    The weights were initialized by the \textit{Xavier} initializer and trained using the \textit{backpropagation through time} algorithm with the \textit{Adam} optimizer \cite{xavier2010init, williams1990efficient, diederik2014adam}.
    The learning rate was set to 0.02 for the first 10 epochs, 0.01 for the next 10 epochs, and 0.005 for the remaining epochs.
 
	\begin{figure}[!t]
	\begin{minipage}[t]{.99\linewidth}
	\centerline{\epsfig{figure=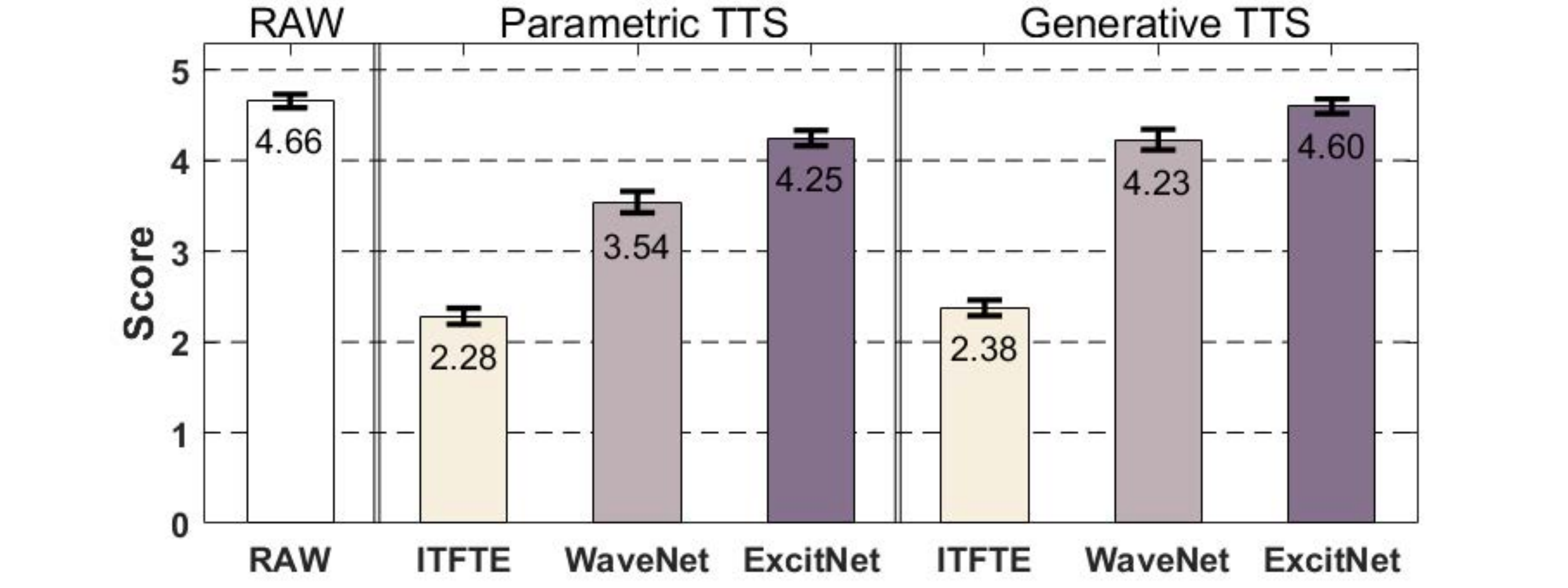,width=90mm}}
	\vspace*{-2pt}	
	\centerline{(a)}  \medskip
	\vspace*{-6pt}	
	\end{minipage}
	\begin{minipage}[t]{.99\linewidth}
	\centerline{\epsfig{figure=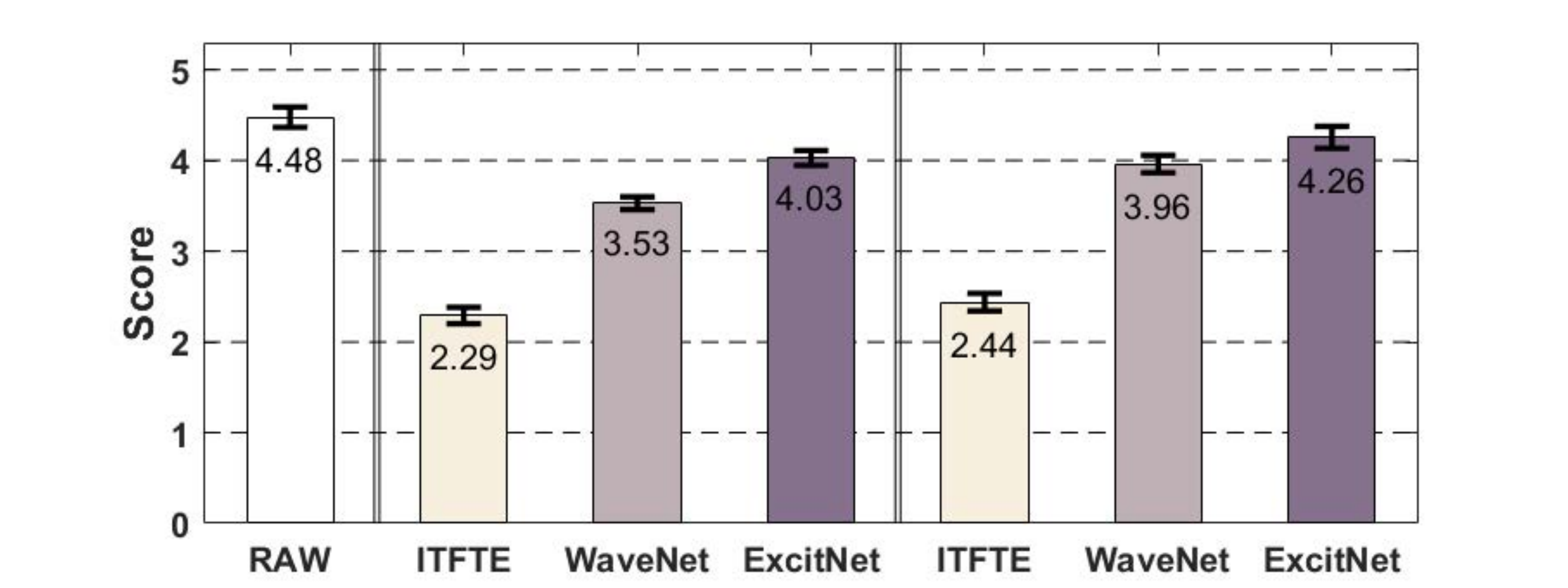,width=90mm}}
	\vspace*{-2pt}	
	\centerline{(b)}  \medskip
	\end{minipage}	
	\vspace*{-6pt} 
	\caption{
	MOS results with 95\% confidence intervals. Generated acoustic features were used to compose the input auxiliary features for the different vocoding models: (a) Korean female and (b) Korean male speakers.}	
	\label{fig:mos}
	\end{figure}

\subsubsection{Evaluation results}
\label{ssec:ch5-1-4}

    To evaluate the perceptual quality of the proposed system, mean opinion score (MOS) tests were performed\footnote{Generated audio samples are available at the following URL:\\ \url{https://soundcloud.com/eunwoo-song-532743299/sets/tacotron2_excitnet}}.
    Eighteen native Korean speakers were asked to make quality judgments about the synthesized speech samples, using the following five possible responses: 1 = Bad; 2 = Poor; 3 = Fair; 4 = Good; and 5 = Excellent.
    In total, 40 utterances were randomly selected from the test set and were then synthesized using the different generation models.
    In particular, to confirm performance differences between vocoding models, speech samples synthesized by both the parametric ITFTE vocoder and the conventional WaveNet vocoder with both the conventional parametric TTS model and the seq2seq-based generative TTS model were also evaluated together.

    Figure~\ref{fig:mos} shows the MOS test results with respect to different acoustic parameter estimation models, and the analysis can be summarized as follows: (1) The seq2seq model performed better than the conventional parameter estimation approach did, regardless of the vocoding model type. 
    This implies that the seq2seq model is effective for modeling spectral and excitation parameters, which results in improved perceptual quality of synthesized speech. 
    (2) When analyzing the results only with the vocoding method, the ExcitNet vocoder performed significantly better than other vocoders did not only in the conventional parameter estimation method but also in seq2seq-based frameworks.
    Because the ExcitNet directly generates time-domain excitation signals, the smoothing effect of acoustic features generated by both acoustic models could be decreased. 
    (3) Consequently, the seq2seq acoustic model with the ExcitNet vocoder achieved 4.60 and 4.26 MOS results for the Korean female and male speakers, respectively.

	\begin{table}[!t]   
	\begin{center}         
	\caption{Number of utterances in different sets and emotions for the Korean female speaker.}  
	\vspace*{-8pt}
	\label{table:numUttEmo}
	{\small        
	\begin{tabular}{>{\centering}m{.145\linewidth}||c|c|c}
	\Xhline{2\arrayrulewidth}
	Emotion			& Training  & Development 	& Test \\
			\hline \hline
	Neutral		& 4,976 (10.0 h)		& 280 (0.5 h)			& -	\\
	Happy		& 3,710 (3.7 h)		& 500 (0.5 h)			& 500 (0.5 h)	\\
	Sad		& 3,680 (3.7 h)		& 490 (0.5 h)			& 490 (0.5 h)	\\
			\hline 
	Total 	& 12,366 (17.4 h)		& 1,270 (1.5 h) 			& 990 (1.0 h)	\\
			\Xhline{2\arrayrulewidth}
	\end{tabular}}          
	\end{center}         
	\vspace*{-12pt}
	\end{table}

\subsection{GST-Tacotron 2 with ExcitNet vocoder for expressive speech synthesis}
\label{ssec:ch5-2}

    To verify the effectiveness of the proposed method as an expressive speech synthesis system, we combined the GST-based emotional style embedding network with the Tacotron 2 model. 
    The GST-Tacotron 2 has shown a capability to extract a high-dimensional embedding that implicitly contains the speaker's prosody and style information, and the ExcitNet has performed robustly when producing prosody-modified speech segments \cite{kwon2019emo, song2018speaker}.
    Thus, combining these two models was expected to generate more natural speech signals when compared with the conventional approaches.
    
\subsubsection{Database}
\label{ssec:ch5-2-1}    

    In addition to the Korean female speaker's speech database as described in section~\ref{ssec:ch5-1-1}, we prepared expressive speech corpus that contained happy and sad emotions recorded by the same speaker.
    Table~\ref{table:numUttEmo} shows the number of utterances used in the experiments.
    
\subsubsection{GST-based emotional style embedding model}
\label{ssec:ch5-2-2}        

    To train the GST-Tacotron 2 model, the acoustic features were used as an input of the GST-based network to learn emotional characteristics of speech corpus. 
    A 256-dimensional emotion embedding extracted from the GST network was duplicated as the length of transcript embedding, and it was concatenated to the transcript embedding to generate emotion-related acoustic features via the seq2seq decoder network.
    Note that the reference encoder had six stacked 2-D convolution layers and a uni-directional GRU layer with 128 units.
    The six convolution layers had 32, 32, 64, 64, 128, and 128 output channels; in addition, a 4-headed attention network and 10 GSTs were used to construct the style token layer.
    The training criterion and learning rate schedule were the same with the setup for the seq2seq acoustic model, as mentioned in section~\ref{ssec:ch5-1-3}.

    In the inference phase, average weight vectors were first obtained from the training set via the centroid of the cluster composed of all the emotion embedding per emotion, and then the proposed system was given the text-weight matrix pairs to generate emotional speech \cite{kwon2019emo}.

	\begin{table}[!t]   
	\begin{center}         
	\caption{
	MOS results with 95\% confidence intervals for the expressive speech synthesis system with respect to the different vocoding models: Acoustic features generated from the seq2seq model were used to compose the input auxiliary features. The systems that returned the best MOS quality are in a bold font.}
	\vspace*{-3pt}
	\label{table:mos_emo}
	{\small        
	\begin{tabular}{>{\centering}m{.20\linewidth}||c|c}
	\Xhline{2\arrayrulewidth}
	System	    & Happy & Sad \\
			\hline \hline
	Raw         & 4.72$\pm$0.06 & 4.65$\pm$0.07 \\\hline
	ITFTE       & 2.61$\pm$0.13 & 2.52$\pm$0.12 \\
	WaveNet     & 4.27$\pm$0.08 & 4.09$\pm$0.12 \\
	ExcitNet    & \textbf{4.67$\pm$0.05} & \textbf{4.43$\pm$0.09} \\
			\Xhline{2\arrayrulewidth}
	\end{tabular}}          
	\end{center}         
	\vspace*{-12pt}
	\end{table}

\subsubsection{Evaluation results}
\label{ssec:ch5-2-3}

    To evaluate the quality of the expressive speech synthesis system, the MOS tests were performed\footnote{Generated audio samples (expressive versions) are available at the following URL:\\\url{https://soundcloud.com/eunwoo-song-532743299/sets/gst_tacotron2_excitnet}}.
    The test setups were the same as those described in section~\ref{ssec:ch5-1-4}, except that 20 happy and 20 sad utterances were randomly selected from the test set and were then synthesized using the different vocoding models with the same seq2seq acoustic model.

    Table~\ref{table:mos_emo} shows the MOS test results with respect to different generation models.
    The results on the perceptual quality of emotional speech vocoded by the ExcitNet were significantly better than those of the two conventional vocoders, regardless of the speaker's emotional state.
    In other words, the effectiveness of the ExcitNet when producing prosodic speech segments was also demonstrated in the seq2seq acoustic model.
    The proposed system with the ExcitNet vocoder achieved 4.67 and 4.43 MOS results for happy and sad emotions, respectively.
    
\section{Conclusion}

    This paper proposed a high-quality generative TTS speech synthesis system comprising an effective spectrum and excitation estimation method and a generative model.
    By combining the Tacotron 2-based acoustic model as a feature estimator with the ExcitNet-based neural vocoder, the proposed method significantly improved the perceptual quality of synthesized speech.
    The additional merit of the proposed method was confirmed when producing expressive speech segments by adopting a GST-based emotion embedding method.
    The experimental results verified that the proposed systems trained by either narrative or emotional speech corpora performed significantly better than the system with a conventional WaveNet vocoder and our best prior parametric TTS counterpart.

\bibliographystyle{IEEEtran}
\bibliography{mybib}

\end{document}